# Review article

# Nanoelectrodes for Intracellular Measurements of Reactive Oxygen and Nitrogen Species in Single Living Cells

Keke Hu,[1,§] Yan-Ling Liu,[2] Alexander Oleinick,[3] Michael V. Mirkin,[1] Wei-Hua Huang[2] and Christian Amatore[4,3,*]

**Addresses**
[1] Department of Chemistry and Biochemistry, Queens College-CUNY, Flushing, NY 11367, USA
[2] College of Chemistry and Molecular Sciences, Wuhan University, Wuhan, 430072, China
[3] PASTEUR, Departement de chimie, Ecole Normale Superieure, PSL University, Sorbonne University, CNRS, 75005 Paris, France
[4] State Key Laboratory of Physical Chemistry of Solid Surfaces, College of Chemistry and Chemical Engineering, Xiamen University, Xiamen, 361005, China
[§] Present address: Department of Chemistry and Molecular Biology, University of Gothenburg, Kemivägen 10, 412 96 Gothenburg, Sweden

* Corresponding author: Amatore, Christian (christian.amatore@ens.psl.eu)

**Summary:** Reactive oxygen and nitrogen species (ROS and RNS) play important roles in various physiological processes (e.g., phagocytosis) and pathological conditions (e.g., cancer). The primary ROS/RNS, viz., hydrogen peroxide, peroxynitrite ion, nitric oxide, and nitrite ion, can be oxidized at different electrode potentials and therefore detected and quantified by electroanalytical techniques. Nanometer-sized electrochemical probes are especially suitable for measuring ROS/RNS in single cells and cellular organelles. In this article, we survey recent advances in localized measurements of ROS/RNS inside single cells and discuss several methodological issues, including optimization of nanoelectrode geometry, precise positioning of an electrochemical probe inside a cell, and interpretation of electroanalytical data.

## Introduction: Scope and Context of the Problem

Oxidative stress conditions are encountered by all aerobic organisms during their whole life. Indeed, aerobic cells mostly derive their energy from the intracellular enzyme-catalyzed oxidation of fat and sugars to $CO_2$. Also, metalloenzymes which are central actors of the respiratory chain in mitochondria are generally good reducing agents, prone to open side routes leading to $O_2$ reduction to superoxide ion ($O_2^{\bullet-}$) that is the precursor of a series of hazardous species collectively named as "reactive oxygen species (ROS)" and "reactive nitrogen species (RNS)" [1-4]. ROS and RNS may induce molecular damages to almost all organic compounds performing biological functions (nucleic acids, proteins, cells carbohydrates and lipids, etc.) – a situation termed "oxidative stress" when it runs out of control. Even without exposure to radiation or other photo-biological effects, oxidative stress can bring about such pathological conditions as inflammation, carcinogenesis, Parkinson and Alzheimer diseases, and various autoimmune illnesses, as well as accelerated ageing.

Thus, ROS, RNS and oxidative stress are often perceived as harmful. However, this is not correct: oxidative stress also plays many beneficial roles at the level of cells and tissues. Aerobic cells may initiate crucial processes involving oxidative stress to control several positive redox-sensitive mechanisms, e.g., those leading to the apoptotic elimination of impaired cells. ROS and RNS are also essential for phagocytosis in macrophages and neutrophils that eliminates pathogens such as bacteria and viruses. Moreover, NO$^\bullet$ – a major RNS – is a pervasive signaling molecule involved in many physiological and pathological processes; it is also a potent vasodilator and an efficient bactericide. In fact, several aerobic cells may leave their resting state to produce significant quantities of ROS and RNS through the involvement of specific enzymes such as NADPH oxidases and NO-synthases that actively catalyze the production of $O_2^{\bullet-}$ and NO$^\bullet$, respectively [5].

The importance of oxidative stress under "normal" conditions, i.e., when not due to irradiation, has been well recognized since the 90's, and it has subsequently become a subject of many biomedical studies mostly focused on the long-term (months, years for medical concerns) or middle-term (hours, days for metabolites, gene expression) consequences. Nonetheless, detailed studies of the primary stages of intracellular ROS and RNS generation (milliseconds to minutes) have long been hampered by absence of sufficiently sensitive analytical methods for measurement inside living cells under physiological conditions.

Due to the electroactivity of most primary ROS and RNS, electrochemistry at platinized ultramicroelectrodes (Pt@UME) has been extensively employed in studies of oxidative stress [6]. However, they severely affected metabolic functions of the investigated cells upon penetrating through cellular membranes because of their relatively large size [7]. Hence, measurements of ROS and RNS were mostly limited to the detection of ROS and RNS in the near vicinity of cell surfaces by analyzing the fluxes released from single cells [6, 8]. Although indirect and extracellular, these studies yielded some significant results, including the first evidence that primary oxidative stress bursts consist of a cocktail of $H_2O_2$, peroxynitrite ions (ONOO$^-$), NO$^\bullet$ and nitrite ions ($NO_2^-$) formed by the rapid parallel- or cross-evolution of $O_2^{\bullet-}$ and NO$^\bullet$ [6, 8].

Nanoelectrodes have been designed over the past 15 years [9] and used for investigating many biological topics such as mapping "hot spots" on plasma membranes [10, 11], intrasynaptic activity [12, 13] and other intracellular events [14-17]. Their conical shape gave these nanoelectrodes a sharp and strong tip [10] but prevented them from being deeply inserted inside cells because of potential major damage to the membrane [7]. Nanoelectrodes with near-cylindrical sheath may enable resealing of the cell membrane around the electrochemically inert shaft after the penetration. This would minimize the damage from electrode insertion into the cytoplasm and allow the nanoelectrode tip to be positioned where the measurement is desired. Nanotubes [18, 19, 20], nanowires of tungsten [21] or noble metals [22, 23] have been proposed but they generally lack the properties allowing their routine fabrication. Conversely, when these probes were adequate, their size and/or electrochemical activity were not suitable. The development of nanoelectrodes appropriate for investigating intracellular mechanisms without significantly altering the cellular behaviors has remained an important challenge up to the recent years. In the following we wish to illustrate the success of a few distinct approaches developed in the laboratories of the authors for providing a better understanding of oxidative stress at the single cell level.

## Possible nanoelectrode geometries

As discussed above, only nanoelectrodes with near-cylindrical insulating shafts and electroactive tips can be used to detect the primary ROS and RNS (viz., $H_2O_2$, $ONOO^-$, $NO^\bullet$, $NO_2^-$) [6] fluxes and concentrations localized within nanoscale domains inside single cell cytoplasms.

**Figure 1**

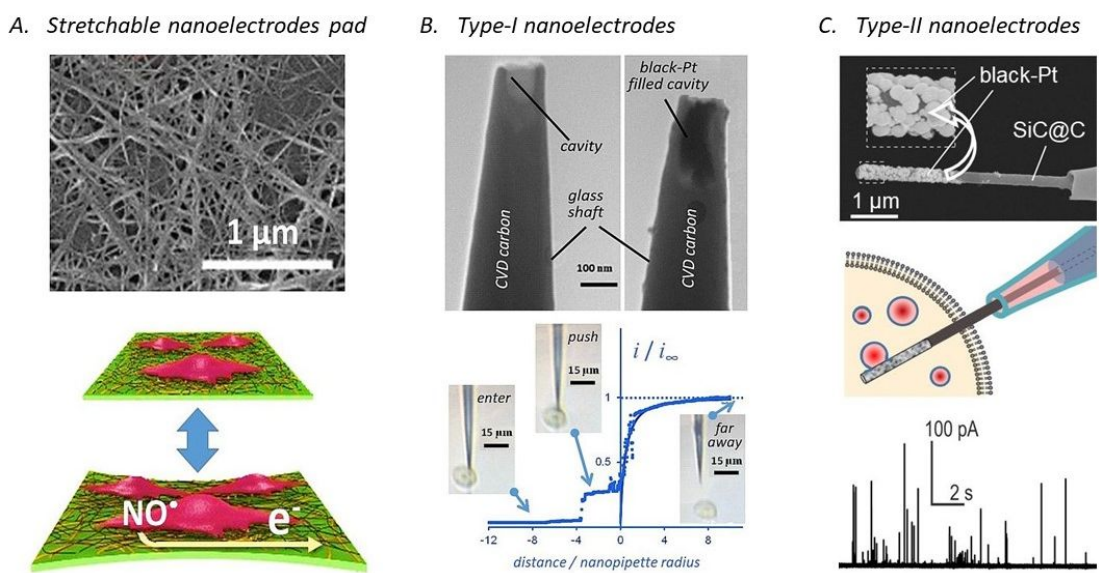

Representation of the three types of nanoelectrodes discussed in this review. (A) Stretchable pad electrode consisting of gold nanotubes (Ø 60 nm) interlaced with carbon nanowires (Ø 2 nm) on a PDMS thin pad [30]; top: SEM image; bottom: schematic view of cells cultured on a pad before and after stretching. (B) Top: Type-I nanoelectrode; TEM of a glass nanopipette filled with chemical vapor deposition (CVD) of carbon before and after Pt black was deposited in the cavity; bottom: SECM-type approach curve recorded with a Type-I nanoelectrode tip (Ø 160 nm) moving toward the membrane of a MCF-10A human breast cell, as shown in the optical micrographs [37]. (C) Type-II nanoelectrode; top: SEM view of the nanoelectrode tip consisting of a SiC nanowire (Ø 200 nm) modified by a 50 nm carbon CVD film onto which Pt black (shown in gray shade) has been electrodeposited; middle: sketch of the platinized SiC@C nanowire inserted inside a RAW 264.7 cell and probing one phagolysosome [39,40]; bottom: partial trace of current spikes featuring ROS/RNS oxidation when a series of individual phagolysosomes collapse onto the Pt black tip.

Yet, before discussing these nanoelectrodes in more details we wish to stress that ROS and RNS may act as important signaling messengers emitted by cells in their close vicinity. This is well recognized for $NO^\bullet$ and its vasodilatory role in the cardiovascular system [24], yet ROS and RNS are also thought to act as primary biochemical signals initiating mechanotransduction cascades allowing cells to adapt to mechanical stresses [25, 26]. The first attempts to investigate mechanotransduction with electrochemical sensors were performed by positioning rigid Pt@UMEs over single cells [27, 28]. These approaches opened the field to electrochemical investigations and provided significant data inaccessible by other analytical techniques. However, their too rigid configurations did not allow stressing and probing cells under conditions similar to natural ones. Stretchable electrodes consisting of felts of hybrid nanowires interlaced onto PDMS films were developed to overcome this important limitation [29-31]. Such deformable pads allowed stressing mechanically different types of human cells adhering onto the sensor surfaces while monitoring their responses (Figure 1A). This permitted to establish for the first time that brief signals consisting of NO• spikes of intensities correlating with the stress intensity were instantly released by stressed cells when L-arginine was available and NO-synthase pools (eNOS) not inhibited. Conversely, when eNOS were inhibited or the cells overstretched, another ROS was released, most probably $H_2O_2$ [32]. Strictly speaking, such electrochemical sensors stand outside the scope of this review because ROS/RNS were monitored in close contact with cells membranes but nonetheless in the extracellular space. Even so, we wished to mention their existence since on the one hand their high-performance properties and soft material nature are expected to endow them with a wide range of applications for the investigation of molecular messengers emitted by cells; and on the other hand, the mechanism of cell transduction can be further demonstrated by intracellular monitoring of ROS and RNS with nanoelectrodes as presented below.

To monitor and characterize the primary ROS/RNS inside aerobic living cells, i.e., cells placed in aerated solutions, one cannot rely on cathodic voltammetric measurements since $O_2$ (present at ca. 0.23 mM) reduction will mask cathodic waves of ROS/RNS present in much lesser concentrations [33]. Interestingly, the $H_2O_2$, $ONOO^-$, NO• and $NO_2^-$ oxidation waves at Pt black electrodes are sufficiently separated for allowing each species dynamic flux or concentration to be monitored with sufficient accuracy and sensitivity [8]. Furthermore, Pt-black tips electrodeposited by reduction of a Pt salt, e.g., $H_2PtCl_6$, prevent any rapid poisoning of the electroactive surfaces while oxidizing $H_2O_2$ and/or $NO_2^-$ [34, 35].

To the best of our knowledge only two main designs simultaneously fit these specifications. Type-I is a nanopipette almost filled up by a conductive material, e.g., chemical vapor deposited (CVD) carbon [36], with a nanocavity filled with Pt black [16, 37-38]. Such a design allows exposing a Pt black disk protected by a thin glass rim to the external solution (Figure 1B). The second design (Type-II) relies on commercial mechanically robust nanorods to produce the required cylindrical shaft. SiC semiconductor nanowires become suitable for amperometric experiments after being coated with a thin CVD carbon film to remedy their poor electrical conductivity. Electrodepositing nanometer-thick sleeves of Pt black over a micrometric length at their tips produced robust ROS/RNS suitable sensors (Figure 1C) [39, 40].

**Biological applications**

In this section we discuss a few representative applications of nanoelectrodes to intracellular monitoring of ROS/RNS that could not be attained by other analytical methods (see, e.g., refs [6, 16, 30, 31, 32, 37, 38, 39, 40] for discussions of this important point under many biological situations). Type I nanoelectrodes were used to compare the primary ROS/RNS contents in normal (MCF-10A), cancerous (MDA-MB-231) and metastatic (MDA-MB-468) human breast cells [37]. In cancerous cells, large quantities of ROS/RNS were released through a series of sequential bursts. By contrast, ROS/RNS concentrations in normal cells were lower than the detection limit. Nonetheless, intense intracellular ROS/RNS bursts similar to those displayed by metastatic cells were observed after treating normal cells with diacylglycerol-lactone [37] (Figure 2A) to activate protein kinases C [41], unraveling the complexity of the connection between metastatic features and oxidative stress, and hence stressing the need for the introduction of nano-analytical methods such as this one to evaluate new prospective targets for cancer therapy.

**Figure 2**

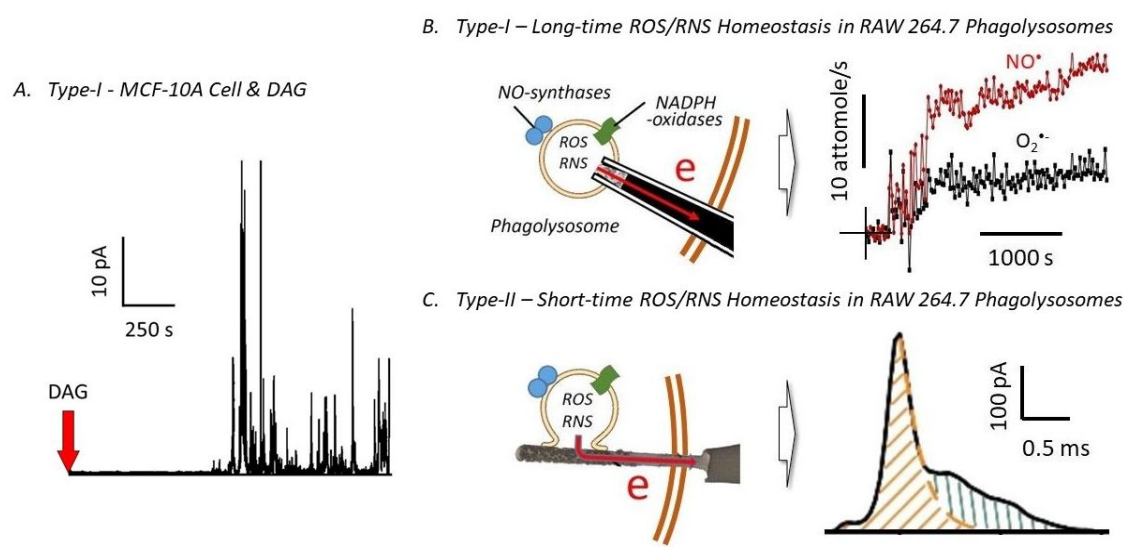

Biological applications of Type-I and Type-II nanoelectrodes. (A) Monitoring of current spikes featuring intracellular oxidative stress bursts when a normal (MCF-10A) human breast cell is submitted to diacylglycerol-lactone (DAG); the vertical red arrow indicates the moment when DAG was added to the cell culture to activate protein kinases C [37]. (B,C) Investigations of ROS/RNS homeostasis in phagolysosomes of murine macrophages RAW 264.7 activated by interferon-γ and LPS; (B) long-time characteristics as monitored with a Type-I nanoelectrode inserted within a tightly sealed phagolysosome membrane analyzed in terms of the resulting fluxes of NO$^\bullet$ and O$_2^{\bullet-}$ [38]; (C) short-time behavior observed after collapse of one phagolysosome onto the Pt black tip of a Type-II nanoelectrode: the yellow-hatched area of the current spike represents the oxidation current of ROS and RNS initially contained in the phagolysosome, while the current component due to the ROS and RNS generated after its collapse onto the nanoelectrode is represented by the gray-hatched area [40].

This example was focused on cancer, i.e., a negative outcome of oxidative stress. We wish now to address a less known but extremely positive role of oxidative stress, i.e. phagocytosis – an essential feature of the immune system performed by macrophages within their intracellular vacuoles (phagolysosomes) to eliminate impaired cells and pathogens. Based on extracellular detection [32, 42], it was established that phagocytosis involves a cocktail of H$_2$O$_2$, ONOO$^-$, NO$^\bullet$ and NO$_2^-$ species formed by the fast chemical evolution of O$_2^{\bullet-}$ and NO$^\bullet$ generated within phagolysosomes by activated NAD(P)H oxidases and NOS-synthases. Nonetheless, two important issues could not be addressed in those studies, one of which is possible diffusion of some of these reagents (particularly the apolar protonated peroxynitrite *trans*-conformer [43]) across phagolysosomes membranes and their accumulation inside the cells cytoplasm. Type-I platinized nanoelectrodes were used to detect ROS/RNS inside RAW 264.7 cells activated by interferon-γ and LPS and detect ROS and/or RNS. This result suggests that very small quantities of ROS/RNS could leak out from intracellular phagolysosomes [16]; yet these species disappeared within a few seconds and, in any case, their amounts were considerably less than those released from phagolysosomes to the extracellular fluid and monitored with Pt@UME [42]. This experiment demonstrated for the first time that some ROS/RNS could spill out from phagolysosomes and that macrophages are efficiently equipped to rapidly eradicate them. Although obviously important for our immune system and our health, this process remained a subject of speculations [44] until ROS and RNS could be detected by their oxidation within the cytoplasm of single living cells.

Another issue of high significance for our immune system is the possibility of the ROS/RNS homeostasis inside phagolysosomes. When cell debris or pathogens are captured by macrophages and sequestrated inside phagolysosomes to be degraded, it is essential that the quantities of generated ROS and RNS are sufficient to complete the task but not far excessive to avoid their spilling outside and damaging the macrophage inner components (see above). On the other hand, it is evident that macrophages cannot evaluate the required ROS/RNS quantities when they engulf the cell debris or pathogens. This observation suggests the existence of yet unknown homeostatic mechanisms [45-47] allowing phagolysosomes to continuously produce ROS/RNS and maintain them at safe levels for the macrophage but sufficient for "digesting" the captured materials. To investigate this issue, smaller Type-I platinized nanoelectrodes were inserted inside phagolysosomes of interferon-γ and LPS activated RAW 264.7 and used to monitor their ROS/RNS contents over about one hour [38]. After a variable initiation period (ca. ten minutes) during which only the bactericidal NO• species could be detected, an intense production of highly reactive peroxynitrite ions [48] started and was sustained at a near constant level over the measurement period (Figure 2B). This pointed for the first time to the existence of a ROS/RNS homeostatic mechanism inside a phagolysosome. Notwithstanding the high significance of the above results, only a few sufficiently large vacuoles could be examined because of difficulties in penetrating smaller phagolysosomes while maintaining their integrity over a long time.

This limitation necessitated the development of a complementary strategy allowing rapid monitoring of ROS/RNS inside large numbers of phagolysosomes of widely different sizes. This approach relies on introducing a Type-II platinized nanowire into a living macrophage and monitoring phagolysosomes collapsing onto its Pt black tip and releasing their ROS/RNS contents [17, 49-50]. The main advantage of this method is that a considerable number of transient spikes featuring individual collision events can be collected and analyzed (Figure 1C), thus yielding a wealth of statistically significant data encompassing smaller phagolysosomes whose size would not allow a Type-I nanoelectrode tip to be brought inside [39, 40]. Its obvious disadvantage is that the duration of one phagolysosome collapse onto the tip of a type-II nanoelectrode is too short to allow (*i*) differentiating ROS and RNS as could be done when using type-I nanoelectrodes, or (*ii*) investigating the homeostatic mechanism over durations exceeding a few milliseconds owing to the rapid loss of biological integrity of the ripped vacuoles (Figure 2C). Nonetheless, the collision-based approach allowed analyzing without any ambiguity the short time efficiency of ROS/RNS homeostasis in most analyzed phagolysosomes [40].

## Conclusion

This short review evidenced that the ability of monitoring and quantifying intracellular time-dependent fluxes and concentrations of ROS & RNS with Types-I & II nanoelectrodes has already enabled scrutinizing important biological phenomena that would have remained speculative or poorly understood without proper analytical tools offering a sufficient chemical, kinetic and spatial resolution. It is certain that many further interesting developments will occur upon adapting these approaches to include resistive-pulse sensing functionalities [51, 52], or when combining them with stretchable electrodes to investigate the nature of causal relationships between intracellular redox status and signaling.

## Acknowledgments

In Paris, this work was supported in parts by CNRS, ENS, Sorbonne University and PSL University (UMR 8640 PASTEUR). In Wuhan, it was supported by the National Natural Science Foundation of China (Grants 21725504, 21675121, and 21804101). MVM acknowledges NSF support of his group's research on nanoelectrochemistry of ROS/RNS (CHE-1763337). CA and MVM acknowledge the support from ANR-NSF bilateral program (ANR-AAP-CE06). CA and WHH acknowledge the LIA CNRS NanoBioCatEchem for its support. CA acknowledges Xiamen University for his position of Distinguished Visiting Professor.